\documentclass[twocolumn]{aastex62}

\newcommand{\he}[1] {He\,{\sc #1}}

\newcommand{\ha}{H$\alpha$}
\newcommand{\hb}{H$\beta$}

\newcommand{\hei}{He~{\sc i}}
\newcommand{\heii}{He~{\sc ii}}

\def\kms{\mbox{${\rm km}\:{\rm s}^{-1}\:$}}

\def\lesssim{\mathrel{\hbox{\rlap{\hbox{\lower4pt\hbox{$\sim$}}}\hbox{$<$}}}}

\def\gtrsim{\mathrel{\hbox{\rlap{\hbox{\lower4pt\hbox{$\sim$}}}\hbox{$>$}}}}

\received{\today}
\submitjournal{ApJ Letters}

\shorttitle{The changing-look wind of Swift~J1858.6-0814}
\shortauthors{Mu\~noz-Darias et al. }

\begin{document}

\title{The Changing-look Optical Wind of the Flaring X-ray Transient Swift~J1858.6-0814}

\correspondingauthor{Teo~Mu\~noz-Darias}
\email{teo.munoz-darias@iac.es}

\author[0000-0002-3348-4035]{T.~Mu\~noz-Darias}
\affiliation{Instituto de Astrof\'isica de Canarias, 38205 La Laguna, Tenerife, Spain}
\affiliation{Departamento de Astrof\'\i{}sica, Universidad de La Laguna, E-38206 La Laguna, Tenerife, Spain}

\author{M. Armas Padilla}
\affiliation{Instituto de Astrof\'isica de Canarias, 38205 La Laguna, Tenerife, Spain}
\affiliation{Departamento de Astrof\'\i{}sica, Universidad de La Laguna, E-38206 La Laguna, Tenerife, Spain}

\author{F. Jim\'enez-Ibarra}
\affiliation{Instituto de Astrof\'isica de Canarias, 38205 La Laguna, Tenerife, Spain}
\affiliation{Departamento de Astrof\'\i{}sica, Universidad de La Laguna, E-38206 La Laguna, Tenerife, Spain}

\author{G. Panizo-Espinar}
\affiliation{Instituto de Astrof\'isica de Canarias, 38205 La Laguna, Tenerife, Spain}
\affiliation{Departamento de Astrof\'\i{}sica, Universidad de La Laguna, E-38206 La Laguna, Tenerife, Spain}

\author{J.~Casares}
\affiliation{Instituto de Astrof\'isica de Canarias, 38205 La Laguna, Tenerife, Spain}
\affiliation{Departamento de Astrof\'\i{}sica, Universidad de La Laguna, E-38206 La Laguna, Tenerife, Spain}

\author{D.~Altamirano}
\affiliation{Physics \& Astronomy, University of Southampton, Southampton, Hampshire SO17 1BJ, UK}

\author{D.~J.~K. Buisson}
\affiliation{Physics \& Astronomy, University of Southampton, Southampton, Hampshire SO17 1BJ, UK}

\author{N.~Castro Segura }
\affiliation{Physics \& Astronomy, University of Southampton, Southampton, Hampshire SO17 1BJ, UK}

\author{V.~A.~C\'uneo}
\affiliation{Instituto de Astrof\'isica de Canarias, 38205 La Laguna, Tenerife, Spain}
\affiliation{Departamento de Astrof\'\i{}sica, Universidad de La Laguna, E-38206 La Laguna, Tenerife, Spain}

\author{N.~Degenaar}
\affiliation{Anton Pannekoek Institute for Astronomy, University of Amsterdam, Science Park 904, 1098 XH, Amsterdam, the Netherlands}

\author{F. ~A.~Fogantini} 
\affiliation{Facultad de Ciencias Astron\'omicas y Geof\'{\i}sicas, Universidad Nacional de La Plata, Paseo del Bosque s/n, 1900 La Plata, Argentina}
\affiliation{Instituto Argentino de Radioastronom\'{\i}a (CCT-La Plata, CONICET; CICPBA), C.C. No. 5, 1894 Villa Elisa, Argentina}

\author{C.~Knigge}
\affiliation{Physics \& Astronomy, University of Southampton, Southampton, Hampshire SO17 1BJ, UK}

\author{D. Mata S\'anchez}
\affiliation{Jodrell Bank Centre for Astrophysics, Department of Physics and Astronomy, The University of Manchester, M13 9PL, UK}

\author{M.~ \"{O}zbey Arabaci}
\affiliation{Physics \& Astronomy, University of Southampton, Southampton, Hampshire SO17 1BJ, UK}
\affiliation{Department of Astronomy \& Astrophysics, Atat\"{u}rk University, Erzurum, Turkey}

\author{J.~S\'anchez-Sierras}
\affiliation{Instituto de Astrof\'isica de Canarias, 38205 La Laguna, Tenerife, Spain}
\affiliation{Departamento de Astrof\'\i{}sica, Universidad de La Laguna, E-38206 La Laguna, Tenerife, Spain}

\author{M.~A.~P.~Torres}
\affiliation{Instituto de Astrof\'isica de Canarias, 38205 La Laguna, Tenerife, Spain}
\affiliation{Departamento de Astrof\'\i{}sica, Universidad de La Laguna, E-38206 La Laguna, Tenerife, Spain}
\affiliation{SRON, Netherlands Institute for Space Research, Sorbonnelaan 2, NL-3584 CA Utrecht, the Netherlands}

\author{J. van den Eijnden}
\affiliation{Anton Pannekoek Institute for Astronomy, University of Amsterdam, Science Park 904, 1098 XH, Amsterdam, the Netherlands}

\author{F.~M.~Vincentelli}
\affiliation{Physics \& Astronomy, University of Southampton, Southampton, Hampshire SO17 1BJ, UK}

\begin{abstract}

We present the discovery of an optical accretion disk wind in the X-ray transient Swift~J1858.6-0814. Our 90-spectrum data set, taken with the 10.4m GTC telescope over 8 different epochs and across five months, reveals the presence of conspicuous P-Cyg profiles in \hei\ at 5876 \AA\ and \ha. These features are detected throughout the entire campaign, albeit their intensity and main observational properties are observed to vary on time-scales as short as five minutes. In particular, we observe significant variations in the wind velocity, between a few hundreds and $\sim 2400$ \kms. In agreement with previous reports, our observations are characterised by the presence of frequent flares, although the relation between the continuum flux variability and the presence/absence of wind features is not evident. The reported high activity of the system at radio waves indicates that the optical wind of Swift~J1858.6-0814 is contemporaneous with the radio-jet, as is the case for the handful of X-ray binary transients that have shown so far optical  P-Cyg profiles. Finally, we compare our results with those of other sources showing optical accretion disk winds, with emphasis on V404 Cyg and V4641 Sgr, since they also display strong and variable optical wind features as well as similar flaring behaviour.

\end{abstract}


\section{Introduction}

In addition to a large variety of accretion-related observables, mostly seen in X-rays (\citealt{McClintock2006,vanderKlis2006,Belloni2011}), low-mass X-ray binaries (LMXBs) also show a complex outflow phenomenology. This initially included synchrotron radio emission from jets, either in the form of compact sources or discrete ejections (e.g. \citealt{Mirabel1999}, \citealt{Fender2004}), and subsequently X-ray winds of highly ionized material (e.g. \citealt{Miller2006}, \citealt{Ponti2012}, \citealt{DiazTrigo2016}). These outflows can be the dominant source of power released and mass consumed/expelled by the system during some accretion phases \citep[][]{Fender2016}, and as such represent a fundamental part of the entire accretion process onto stellar-mass black holes (BHs) and neutron stars (NSs).

\begin{table*}
\begin{center}
\caption{Observing log. All the data were taken in 2019.}

\begin{tabular}{clclclc}
Epoch & Observing window (UT) & Grism$^{\dagger}$ and exposures & g-band magnitude\\

\hline
1 & 24/03 (05:14--05:49) & LR ($7\times 280$ s)& 16.1 \\
2 & 14/04 (05:30--06:16) & LR ($9\times 280$ s)& 15.4--15.7\\  
3 & 30/04 (04:32--05:22) & LR ($10\times 280$ s)& 15.9--16.2\\  
4 & 12/05 (03:36--04:32) & LR ($1\times 280$ s) + HR ($10\times280$ s) & 15.7--16.4 \\  
5 & 09/06 (03:52--04:48) & LR ($1\times 280$ s) + HR ($10\times280$ s) & 16.2--16.5 \\
6 & 01/07 (04:03--04:59) & LR ($1\times 280$ s) + HR ($10\times280$ s) & 16.2--16.4\\  
7 & 06/08 (00:00--01:52)) & HR ($21\times 280$ s) + LR ($1\times280$ s) & 16.3 \\
8 & 18/08 (21:57--22:43) & HR ($9\times 284$ s) & 15.7--16.4\\
\hline
\end{tabular}
\end{center}
\vspace{-0.5cm}
\tablenotetext{\dagger}{LR and HR indicate Lower ($\sim 350$ \kms) and Higher ($\sim 160$ \kms) Resolution grism (R1000B and R2500R, respectively)}

\vspace{0.5cm}
\label{log}

\end{table*}

Furthermore, intense and sensitive spectroscopic campaigns carried out over the last few years have unveiled the presence of optical winds in several BH transients. P-Cyg profiles have been discovered in V404 Cyg (\citealt{Munoz-Darias2016,Munoz-Darias2017,MataSanchez2018}; see also \citealt{Casares1991} for detections during the 1989 outburst) and MAXI J1820+070 \citep{Munoz-Darias2019}, while an archival search showed the presence of conspicuous  wind signatures in several outbursts of V4641 Sgr (\citealt{Munoz-Darias2018}; see \citealt{Lindstrom2005} and \citealt{Chaty2003} for earlier reports). In addition, the classical systems GRO~J1655-40 and GX~339-4 showed complex emission line profiles (\citealt{Soria2000, Rahoui2014}, respectively), while the intriguing optical dips of Swift J1357.2-0933 (\citealt{Corral-Santana2013}) have recently been found to be related to disk outflows seen at high orbital inclination (\citealt{Jimenez-Ibarra2019b, Charles2019}). All the above, together with the near-infrared P-Cyg profiles witnessed in at least one luminous NS system (GX~13+1; \citealt{Bandyopadhyay1999}; see also \citealt{Homan2016}), indicate that cold accretion disk winds (i.e. those detected at optical and infrared wavelengths) are a relatively common feature -- perhaps ubiquitous -- in the LMXB accretion phenomena. For the best studied case of V404 Cyg, these are found to have a severe impact on the accretion process and outburst evolution, with an associated mass outflow rate greatly exceeding the accretion rate, albeit this system might represent an extreme case \citep{Munoz-Darias2016, Casares2019}.

Swift~J1858.6-0814 was discovered by the BAT monitor on-board the Neil Gehrels Swift Observatory \citep{Gehrels2004} on Oct 25, 2018, and soon after catalogued as a new galactic X-ray binary transient \citep{Krimm2018}. Since the early phase of the outburst, it became a target of special interest owing to its remarkable flaring behaviour at both X-rays \citep{Ludlam2018} and optical wavelengths \citep{Vasilopoulos2018,Baglio2018a,Paice2018}. This triggered the comparison with the BH transients V404 Cyg and V4641 Sgr by \citet{Ludlam2018}, which was reinforced by telegrams reporting the presence of intrinsic X-ray absorption \citep{Reynolds2018} and optical winds \citep{Munoz-Darias2019b}. In this letter, we present multi-epoch, high-cadence optical spectroscopic observations of Swift~J1858.6-0814 showing  that this system displays optical features indicating the presence of an accretion disk wind.

\section{Observations and data reduction}
We obtained optical spectroscopy using OSIRIS (\citealt{Cepa2000}) attached to the Gran Telescopio Canarias (GTC) at the Observatorio del Roque de los Muchachos in La Palma, Spain. The target was observed in eight different epochs over a time lapse of five months within March-August 2019.  We obtained between 7 and 22 individual spectra per epoch with the grisms R1000B (4200 -- 7400 \AA) and R2500R (5575 -- 7685 \AA) depending on the night (see Table \ref{log}).  This resulted in a total of 90 spectra with a time cadence of $\sim$ five minutes and a velocity resolution of $\sim 350$ or  $\sim 160$ \kms depending on the grism (measured from the FWHM of sky lines; i.e. assuming that resolution is dominated by the 1.0 arcsec slit and not the seeing). Weather conditions were good throughout the campaign, with seeing around $\sim 1$ arsec and clear skies. Only epoch-1 was observed on a bright night at high airmass ($\sim 1.7$) due to the very limited visibility window of the target at that time of the year.  Across the entire campaign the slit was rotated so to include a brighter field star placed $\sim 10$ arcsec South-West from the target. In order to carry out a relative flux calibration, spectra from this object were treated in the same way as those of Swift~J1858.6-0814.  Data were reduced, extracted and wavelength calibrated using IRAF tools, while {\sc molly} and custom {\sc python} routines were used for the analysis. From our g-band acquisition images (typically three per epoch and taken before the spectra) we derive magnitudes in the range 15.4--16.5 across the entire campaign (calibrated against Pan-STARRS).

\begin{figure*}[t]
\epsscale{1.17}
\plottwo{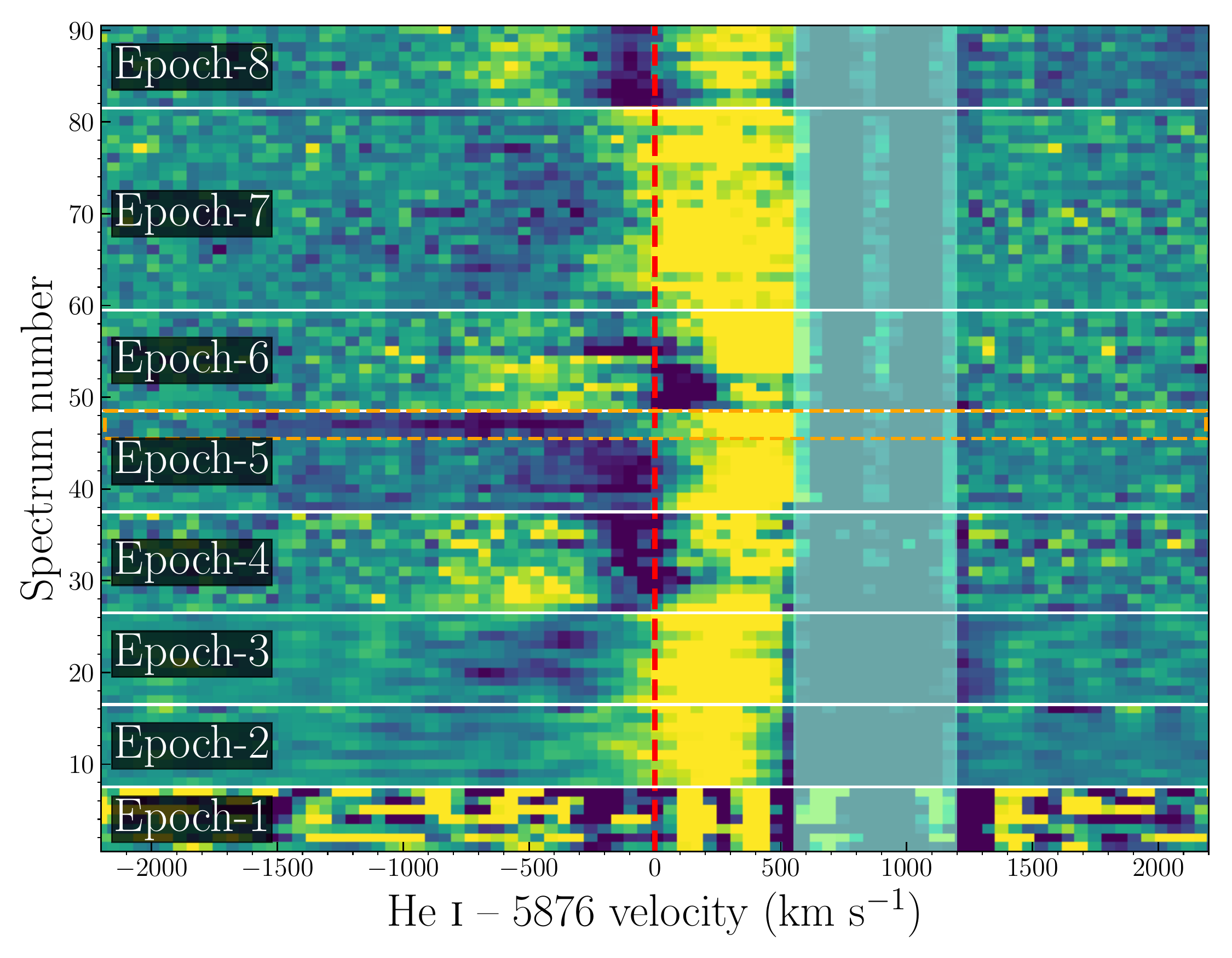}{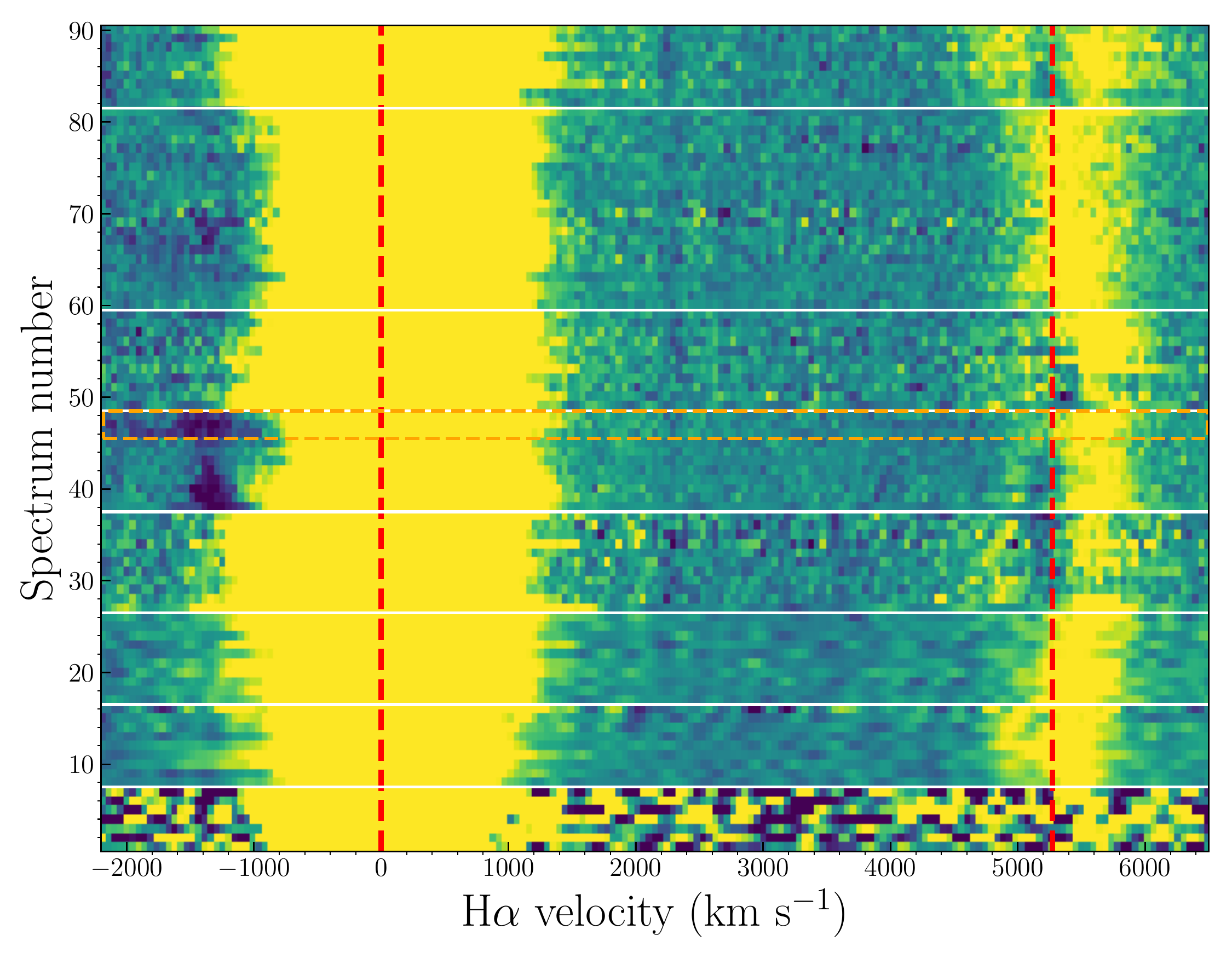}
\plottwo{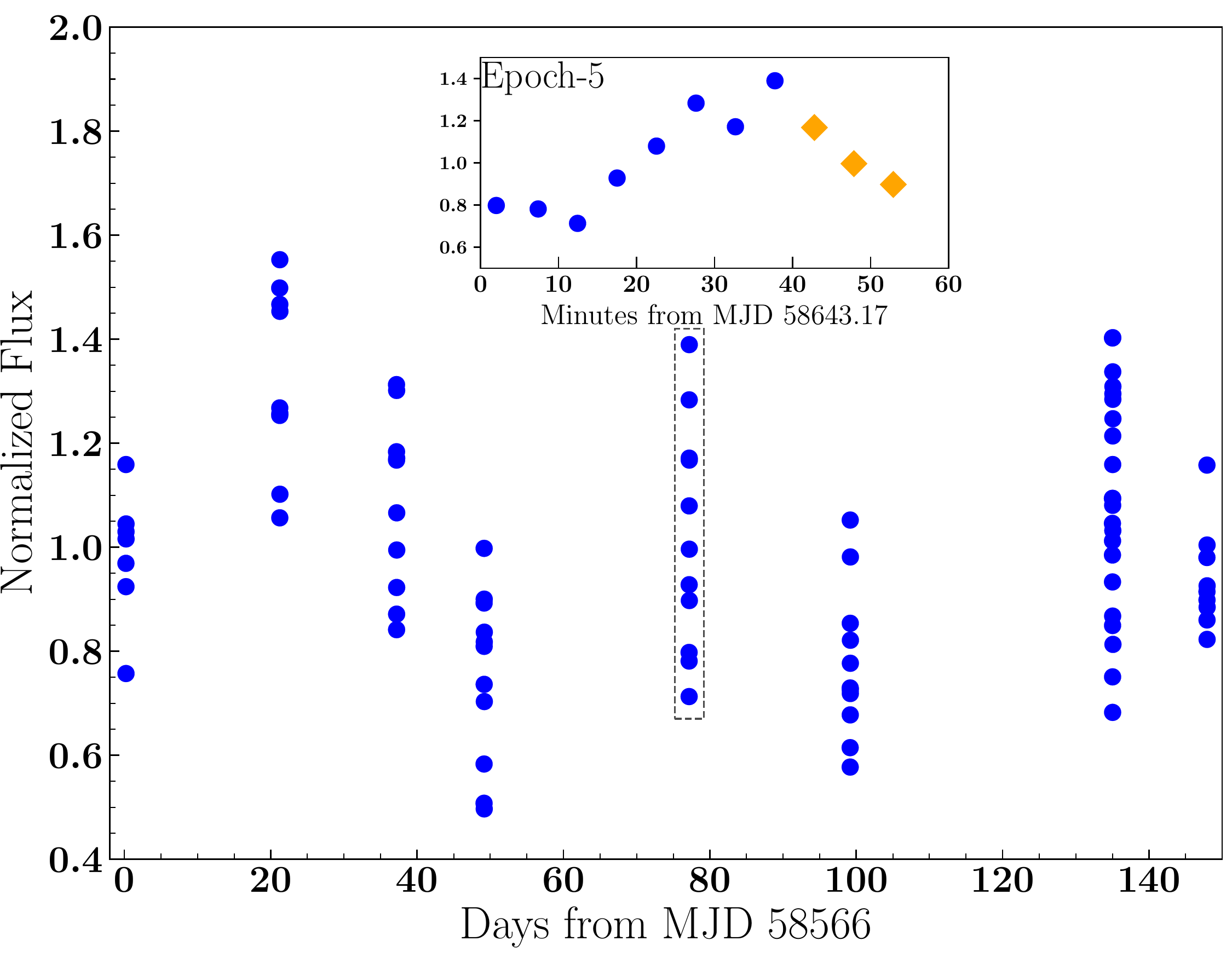}{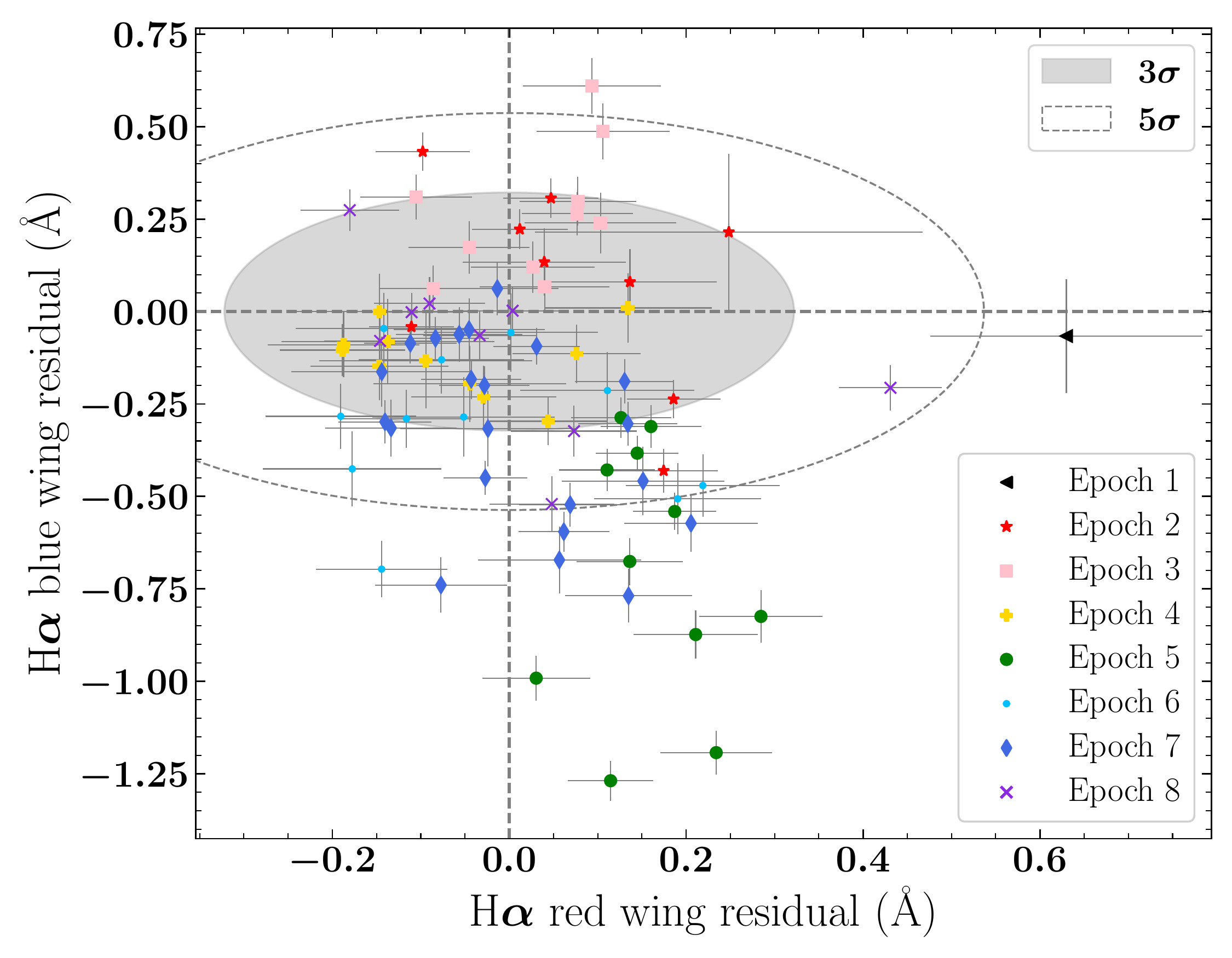}
\caption{Evolution of selected emission lines and the continuum flux across the entire campaign. \textit{Top panels:} trailed spectra for the \he{i}--5876 (left) and \ha\ (+\he{i}--6678; right) spectral regions in velocity scale. Intensity is indicated by a color scale covering from 0.92 (deep blue) to 1.08 (bright yellow) times the continuum level, which was previously normalized. The masked region, containing the Na interstellar doublet in the red wing of \hei-5876, is indicated by a green vertical stripe.  \textit{Bottom-left panel:} evolution of the continuum flux (6000--6250 \AA)  throughout the campaign. Data are represented normalized to the mean value. A zoom-in of epoch-5 (dashed-lined rectangle) is overplotted as an inset, with the orange diamonds corresponding to the spectra with the more conspicuous P-Cyg detections (dashed-lined, orange rectangles in the top panels). \textit{Bottom-right panel:} \ha\ excesses diagram for every spectrum with the exception of epoch-1, for which the epoch-averaged spectrum was used. The residuals are given in equivalent width (\AA). The grey-shaded and dashed-lined circles indicate the $3\sigma$ and $5\sigma$ significance contours, respectively (see section \ref{ewdiagram}).}
\label{fig:multi}
\end{figure*}

\begin{figure*}[t]
\epsscale{1.15}
\plottwo{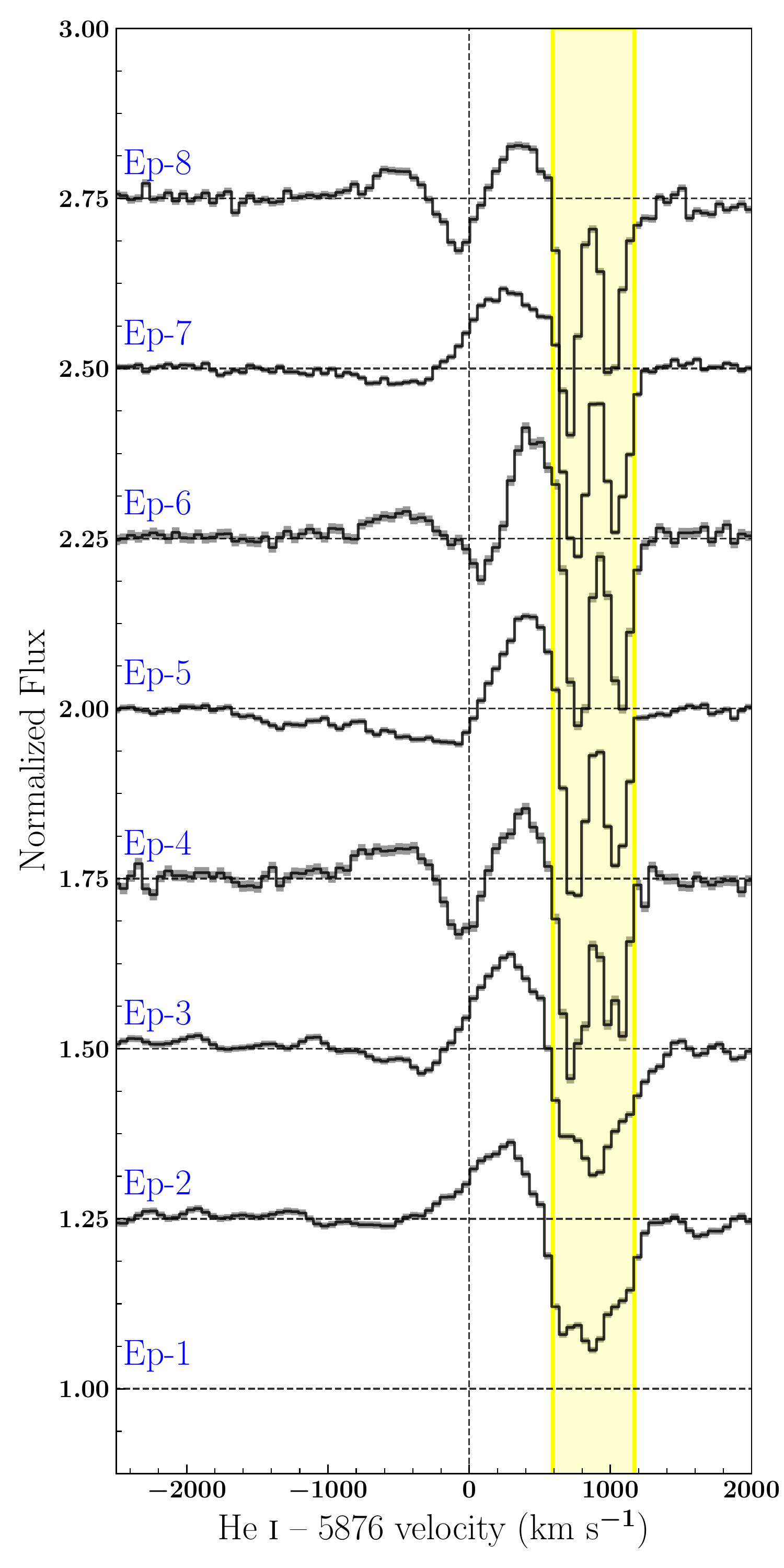}{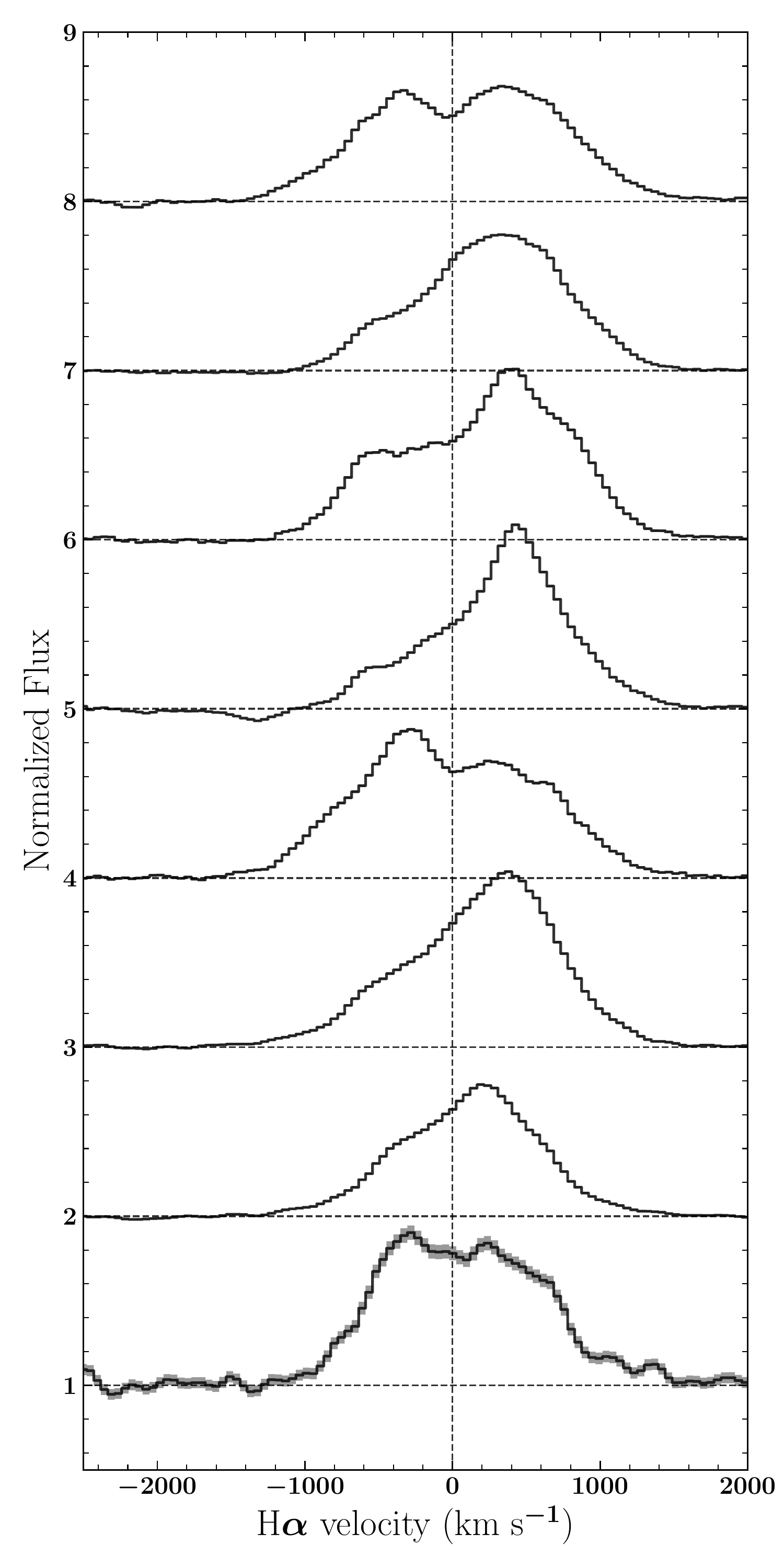}
\caption{Evolution of the \hei-5876 (left panel) and  \ha\ (right panel) emission line profiles using epoch-averaged, normalized spectra. Data are represented using offsets of 0.25 and 1, respectively. The epoch-1  spectrum of \hei-5876 is not represented due to its poor quality. The yellow-shaded region indicates the spectral region contaminated by the Na doublet (interstellar). }
\label{fig:He}
\end{figure*}

\begin{figure*}[t]
\epsscale{1.}
\plotone{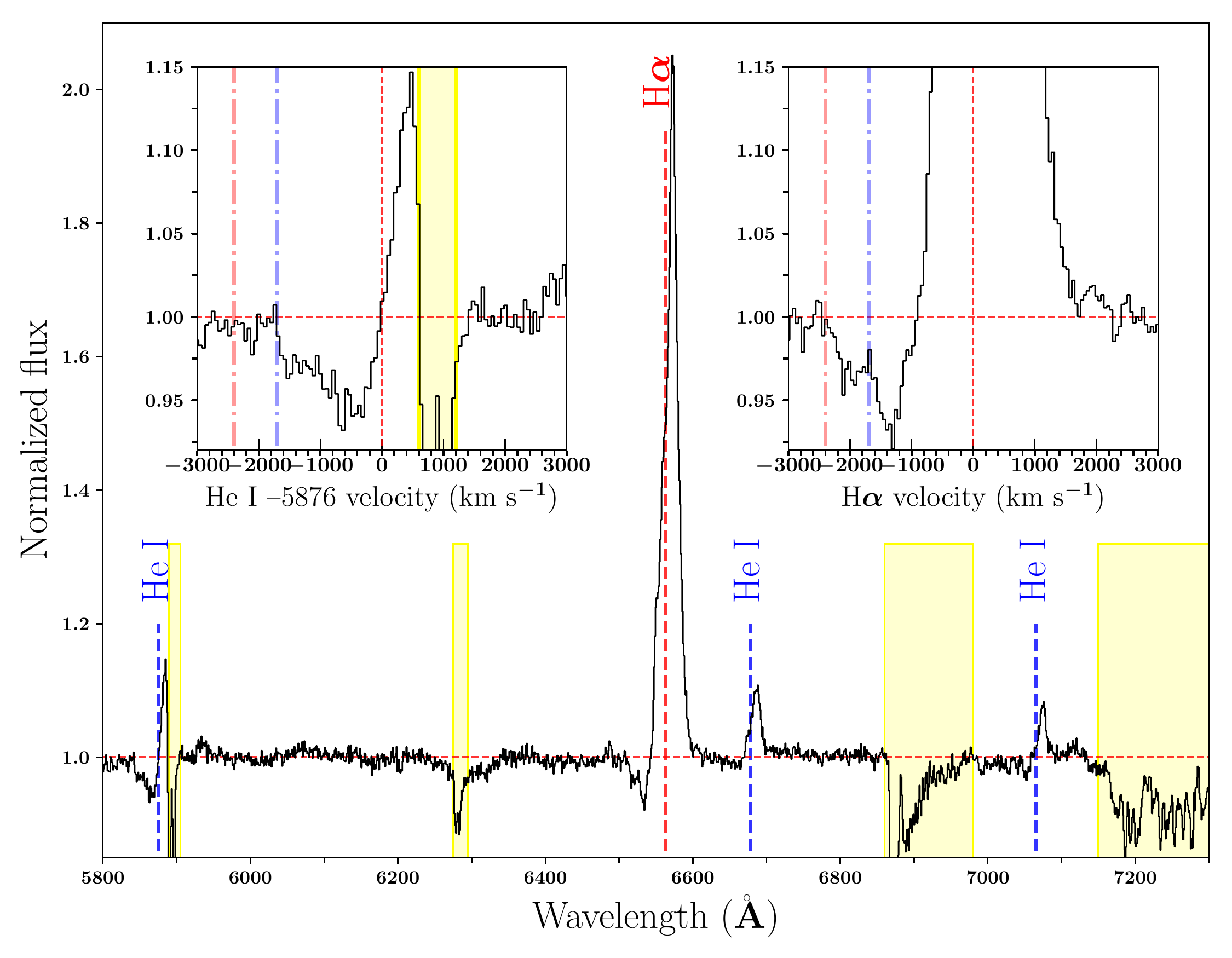}
\caption{Average of the last 3 spectra obtained during epoch-5, when the deepest, high-velocity blue-shifted absorption features are present. The insets show a zoom-in of \hei-5876 (left) and \ha\ (right) in velocity scale, with the blue and red dashed-dotted lines indicating the corresponding blue-edge wind velocities of 1700 \kms and 2400 \kms,  respectively. The yellow-shaded regions indicate spectral regions contaminated by interstellar or telluric features.}
\label{fig:full}
\end{figure*}

\section{Results}
We focus our analysis on the 5500--7200 \AA\ spectral range covered by both the lower (R1000B) and higher resolution (R2500R) grisms. It includes the emission lines  \he{i} at 5876 \AA\ (\he{i}--5876),  \ha\ (6563~\AA), \he{i}--6678, as well as \he{i}--7065. The R1000B spectra (epochs 1-7) also include the Bowen blend (mainly N~\textsc{iii} at 4641 \AA) and \he{ii} at 4686 \AA.  These emission lines are detected across epochs 2 to 7, whilst the limited data quality of epoch-1 only allows to study \ha\ in great detail.

Fig. \ref{fig:multi} shows the trailed spectra (top panels) corresponding to \he{i}--5876 (left) and \ha\ (plus \he{i}--6678; right). Intensity is indicated by a color scale covering from 0.92 (deep blue) to 1.08 (bright yellow) times the continuum level.  Each spectral region was independently normalized by fitting the adjacent, local continuum with a first order polynomial. This was done by considering relatively broad continuum regions at each side of the emission lines ($\sim \pm 10000$ \kms), masking the lines themselves and their closest continuum, as well as other contaminant features (e.g. \he{i}--6678 for \ha). By repeating this analysis with slightly different mask configurations we estimate that the normalization process introduces an uncertainty in the continuum level of $\sim 0.3$ percent (i.e. much lower than the $\pm 8$ percent intensity scale used in the trailed spectra). The blue wing of the \he{i}--5876 emission profile, typically the most sensitive to the presence of optical winds (e.g. \citealt{Munoz-Darias2016}), show blue-shifted absorptions in virtually every observing epoch. These absorptions (represented by deep blue traces) make the blue part of the line disappear completely during a large part of the observing campaign. We interpret this as the signature of an accretion disk wind, similar to those previously witnessed in some BH transients. 

A closer look at the trailed spectra reveals that the blueshifted \he{i}--5876 absorptions are more conspicuous between epochs 3 and 8, whilst \ha\ shows them only on epoch-5 and more weakly on epoch-7 (Fig \ref{fig:multi}, top-right panel). Likewise, \he{i}--6678 behaves in a consistent way with \he{i}--5876, albeit it is significantly less intense than the latter, and therefore any observational feature is expected to be less marked. 

Fig \ref{fig:He} (left panel) shows the evolution of \he{i}--5876 using epoch-2 to -8 nightly average spectra.  We can distinguish two groups of data by looking at the evolution of the blue wing of the emission profile (i.e. negative velocities). Epochs 3, 5 and 7 (and also 2 to some extent) show standard P-Cyg profile shapes (i.e. blue-shifted absorption and red-shifted emission), with the most conspicuous case  (epoch-5)  reaching a terminal velocity\footnote{Throughout the paper, the terminal velocity of the wind is identified as that of the blue-edge of the P-Cyg absorption component. However, we note that this assumption is not always straightforward and is expected to depend on e.g. the physical properties of the ejecta} of $1700$ \kms and an absorption depth at the core of the profile of 95\% the continuum level. An usual approach to determine the terminal velocity in a systematic way is to perform Gaussian fits to the blue-shifted absorption (e.g. \citealt{Munoz-Darias2016}). However, in this case the shapes are clearly non-Gaussian and hence the velocity was simply determined by visual inspection. Given the high signal-to-noise ratio of our data, we estimate this to be accurate within $\sim 100$ \kms (see e.g. insets in Fig. \ref{fig:full}). The second group of data is formed by epochs 4, 6 and 8. They show odd profiles characterized by a strong and slightly blue-shifted absorption troughs reaching down to 92\% the continuum level in some cases (epoch-4 and -8). However, the corresponding epoch-averaged \ha\ lines are dominated by standard double-peaked profiles during these times (Fig \ref{fig:He}; right panel). The combination of a wind-induced blue-shifted absorption at low velocity, together with an underlying  double-peaked profile (observed in \ha) offers a viable explanation for the behaviour of \he{i}--5876 during these epochs. Under this interpretation, the emission bumps at $\sim -500$~\kms would be associated with a partial absorption (by the wind) of the double-peaked disk component. If this is the case, the blue peak would appear weaker than the red one even when \ha\ shows the opposite behaviour (e.g. epoch-4). This dimming of the blue emission suggests that even though the blue edge of the absorption indicates wind velocities of just a few hundreds \kms, a higher velocity component is likely present. 

It is important to remark that while the \ha\ profile seems to be unaffected during these low-velocity wind phases, it is clearly asymmetric or even show blue-shifted absorptions during the remaining epochs. As can be seen  in the trailed spectra (Fig. \ref{fig:multi}), the most clear \ha\ wind detection also occurs on epoch-5. The wind signature is particularly strong in the last three individual spectra of this window (indicated by a dashed-lined, orange rectangle in the trailed spectra), whose averaged spectrum is shown in Fig. \ref{fig:full}. Profound P-Cyg profiles are detected in \he{i}--5876, and especially \ha\, with a blue-shifted absorption reaching 90\% the continuum level and a blue-edge velocity of $\sim$ 2400 \kms. This is larger than $\sim$ 1700 \kms observed in \he{i}--5876 (see insets in Fig. \ref{fig:full}).

\subsection{Continuum flux variability}

Given the strong variability observed in the line spectral profiles, we decided to study the evolution of the continuum flux. This was done by dividing each target spectrum by that of the field star included in the slit and subsequently integrating the flux from 6000 to 6250 \AA. The bottom-left panel of Fig. {\ref{fig:multi}} presents the light-curve, normalized to the mean value. During the campaign, flux is observed to vary within a factor of 3, with significant variability (factor of 2) observed within each epoch. We do not find a strong correlation between the continuum flux and the properties of the wind features found at different epochs. However, there are two facts worth mentioning. On the one hand, epoch-2, the window with the faintest wind features (\he{i}--5876) is the one reaching the highest fluxes. On the other hand, the evolution of epoch-5 (see inset) is affected by the presence of a flare, whose detection is followed by the aforementioned conspicuous wind detections observed in this window (orange dots in the inset). In this regard, we note that even if a similar level of variability is seen in every epoch, this is generally due to fast (non-resolved) flares superposed to a smooth trend, and not to the presence of a single flare as it is the case of epoch-5.   

\subsection{Search for broad emission line wings}
\label{ewdiagram}
Previous studies have shown that, besides P-Cyg profiles, the presence of broad emission line components can be also associated with the presence of winds. Given its strength, \ha\ is the best feature to search for the latter. To this end, we have computed the diagnostic diagram developed in \citet{MataSanchez2018} for V404~Cyg and subsequently refined in \citet{Munoz-Darias2019} for MAXI~J1820+070, to which we refer the reader for further details. We performed a Gaussian fit to the \ha\ line profile masking the innermost part of the line (-500 to 500 \kms in velocity scale), subtracted the fit from the data, and measured the equivalent width (EW) of the residuals in the blue (-2000 to -1000 \kms) and red (1000 to 2000 \kms) emission line wings. Significance levels are computed by measuring the \textit{EW of the continuum}  within masks of the same width (i.e. 1000 \kms) in nearby continuum regions. They show a Gaussian distribution from which significance levels can be derived. The diagram is shown in the bottom-right panel of Fig. \ref{fig:multi}. In order to increase the signal-to-noise ratio, the seven spectra from epoch-1 were averaged and treated as one.  As expected, the entire epoch-5 sits outside the  $3\sigma$ contour, with several spectra located well beyond $5\sigma$, within the P-Cyg region of the plot (i.e. negative blue and positive red residuals). Five spectra from epoch-7 also show significant  blue-shifted absorptions, as well as two from epoch-6.  In addition, there are two spectra sitting on the bottom-left part of the diagram, showing significant blue absorptions ($>5\sigma$). All the above can be considered wind detections based on this method.  However, none of these detections present significant red wing residuals (i.e. $\gtrsim 0.55$), that is, only the blue-shifted part of the P-Cyg is significantly detected. As a matter of fact, the diagnostic diagram do not show any convincing detection of broad emission line wings (top-right region; $>5\sigma$). Only two epoch-3 observations (out of 84) clearly exceed the $3\sigma$ level (again with non-significant red wing residuals), suggesting that broad emission line components are very weak or not present at all. 
 
 \begin{table*}
\begin{center}

\caption{Transient LMXBs with optical P-Cyg profiles }

\begin{tabular}{c c c c c c c}
Transient & Period (h) & Inclination (deg) & P-Cyg depth & Terminal velocity (\kms) & Outburst & References\\

\hline
V404 Cyg & 155.3 & 60--70 & 30\% & 3000 & Non-standard & 1,2 \\
V4641 Sgr  & 67.6 & 60--70 & 40\% &1600 (3000)$^{\dagger}$ & Non-standard & 3,4 \\
Swift~J1858.6-0814 &21.8 (?)& Dipping/eclipsing & 10\% & 2400 & Ongoing & 5, this work    \\
MAXI J1820+070 & 16.9 & 60--81 & 2.5\% & 1800 & Standard & 6,7,8 \\
\hline

\end{tabular}
\end{center}
\vspace{-0.5cm}
\tablenotetext{\dagger}{The maximum P-Cyg terminal velocity is $\sim 1600$ \kms. However, broad wings reaching $\sim$ 3000 \kms were also observed.}
\tablenotetext{}{References: (1) \citet{Casares2014}; (2) \citet{Munoz-Darias2016}; (3) \citet{Orosz2001}; (4) \citet{Munoz-Darias2018}; (5) \citet{Buisson2020c}; (6) \citet{Torres2020}; (7) \citet{Atri2020}; (8) \citet{Munoz-Darias2019}}
\label{sources}
\end{table*} 

\section{Discussion}
\label{discussion}
Swift~J1858.6-0814 is, after the BH systems V404 Cyg, V4641 Sgr and MAXI~J1820+070, the fourth transient LMXB showing optical P-Cyg profiles. \citet{Buisson2020c} have recently reported the detection of several X-ray flares consistent with being type I X-ray bursts. If confirmed, this would make Swift~J1858.6-0814 the first NS transient showing optical wind signatures, reinforcing the similarities between the outflow phenomenology seen in BH and NS transients (e.g. \citealt{Ponti2014} for X-ray winds; \citealt{Miller-Jones2010} for jets). In addition, the system has displayed  profound, periodic dips/eclipses implying a high inclination and strongly suggesting a relatively long orbital period  of 21.8 hours \citep{Buisson2020c}.

As in the previous cases, the emission line that displays the most conspicuous wind signatures is \he{i}-5876. This is also one of the best optical wind markers in massive stars and accreting white dwarfs \citep[e.g.,][]{Prinja1994,Kafka2004}. Our 90 spectra, taken over eight epochs in a time-lapse of  five months, show that the optical wind is active during a large fraction of the time, and continuously detected in our data from (at least) epoch-2 (Figs. \ref{fig:multi} and \ref{fig:He}). Epoch-5 is arguably the most interesting window, with strong wind signatures in both \he{i}-5876 and \ha. The blue-shifted absorptions become particularly conspicuous in the last 3 spectra of this epoch, showing also a significant evolution towards high velocities.  The inset in the bottom-left panel of Fig. \ref{fig:multi} shows the continuum flux evolution during epoch-5. It can be seen that a flare occurs in the middle of the window. Interestingly, the blue-shifted \ha\ absorption almost disappears  during the peak (top-right panel in Fig. \ref{fig:multi}), while the last 3 spectra (orange dots) correspond to the decay of the flare. This behaviour strongly resembles that of V404 Cyg (e.g. fig. 2 in \citealt{Munoz-Darias2016}), which showed the strongest P-Cyg profiles during a low luminosity epoch following a flare. However, the terminal velocity was not observed to vary in that case.  

In V404 Cyg, the presence and variability of the P-Cyg profiles was found to be correlated with the ionisation state of the outer disc, which is traced by the relative strength of \heii\ to \hb\ emission. In particular, the strongest P-Cyg profiles were found at low ionisation; a conclusion that was also supported by the analysis of  MAXI~J1820+070 \citep{Munoz-Darias2019}. Here, we have also computed the flux ratio of \heii\ to \hb\ (as well as the EWs of  \heii\ and Bowen blend) in the 30 spectra taken with the lower resolution grism. Epoch-2 and -3  show variability within a factor of two, roughly following the evolution of the continuum flux, whilst the single value obtained for epoch-5 is in the lower-end of the sample. However, as in the case of the continuum flux variability, we do not find a clear correlation with the presence/absence of wind features.  Nevertheless, given the lack of high-cadence coverage in \heii\ and the similarities between epoch-5 and the behaviour seen in V404 Cyg, we cannot rule out that ionisation effects are a key factor in determining the visibility and properties of the wind. 

\subsection{Comparison with other LMXBs}

Table \ref{sources} displays the main wind observables as well as some fundamental system parameters of the four transient LMXB  with optical P-Cyg detections. As recently discussed in \citet{Hare2020}, the strong flaring activity and variable X-ray absorption of Swift~J1858.6-0814 (see also \citealt{Ludlam2018,Reynolds2018}) resembles the behaviour of V404 Cyg and V4641 Sgr (see e.g. \citealt{Kimura2016,Motta2017b,Morningstar2014, Gallo2014, Munoz-Darias2018}). These two objects display non-standard outbursts characterised by sharp rising phases followed by significant luminosity drops and the absence of steady soft states (e.g. \citealt{Casares2019} for V404 Cyg). The discovery outburst of Swift~J1858.6-0814 is still ongoing and it is beyond the scope of this paper to discuss the outburst evolution. Nevertheless, the source has been active in radio during the entire optical campaign (van den Eijnden et al. to be submitted), which is one of the standard observables associated with BH (and NS to some extent) hard and intermediate states (e.g. \citealt{Fender2016}). This is consistent with the behaviour seen in V404 Cyg, V4641 Sgr and MAXI~J1820+070. 
  
Table \ref{sources} also reports the maximum terminal velocity and blue-absorption depth observed in the four transients with optical P-Cyg detections. These values are likely biassed since (among other things) the observing campaigns (e.g. number and frequency of observations) were significantly different and the wind parameters might also change from outburst to outburst. For instance, the 1989 outburst of V404 Cyg is characterized by lower wind velocities than the 2015 event \citep{MataSanchez2018}. Nevertheless, the numbers suggest that, from a purely observational point of view, the wind signatures of Swift~J1858.6-0814 are reminiscent but not as extreme as those found in V404 Cyg and V4641 Sgr, while they significantly exceed those of MAXI~J1820+070. This last source displayed a regular outburst evolution (e.g. \citealt{Shidatsu2019}) and has an orbital period of  $\sim 17$ hours \citep{Torres2019}. A tempting possibility that arises from Table \ref{sources} is that larger accretion disks (i.e. $\sim$ long orbital periods) might produce stronger winds, that carry away more mass and/or angular momentum. This could impact on the observed outburst evolution, which at least for the cases of V404 Cyg and V4641 Sgr  deviates from the standard patterns typically seen in LMXBs (see e.g. \citealt{Dunn2010, Munoz-Darias2014} for global studies). Clearly, more observations are required to confirm this speculation. A key system on this matter could be GRS~1915+105 -- the BH-LMXB with the longest orbital period \citep[e.g.][]{Corral-Santana2016} -- that have shown some of the best examples of X-ray winds and radio jets (e.g. \citealt{Neilsen2009}). Although this very extinguished source cannot be observed at optical wavelengths, sensitive, infrared spectroscopy might be able to shed light on this topic.

Finally, it is worth discussing the role of the orbital inclination in the wind detectability, since it has been found to be a key parameter for X-ray winds, which are best seen at high inclination \citep{Ponti2012,DiazTrigo2016}. Table \ref{sources} shows how optical P-Cygs have been only detected in sources with relatively high inclination ($\gtrsim$60 deg.; see \citealt{Higginbottom2019} for a discussion on the wind geometry and detectability as a function of the line-of-sight). To this list one could add the BH transient Swift~J1357.2-0933. This high inclination system has not shown standard P-Cyg profiles, but its characteristic optical dips have been recently found to be associated with broad, blue-shifted absorptions indicating the presence of an outflow \citep{Jimenez-Ibarra2019, Charles2019}. 

\section{Conclusions}
We have detected clear optical features indicating the presence of an accretion disk wind in the X-ray transient Swift~J1858.6-0814. The observational properties of the wind are similar to those observed in V404 Cyg and V4641 Sgr. These systems also share other observables with Swift~J1858.6-0814, such as the presence of frequent flares and variable X-ray absorption. As it is the case for other LMXBs with optical wind detections, the outflow is contemporaneous with the radio jet. This work provides additional support for systematic and sensitive optical spectroscopic studies of active X-ray binaries in order to unveil the occurrence rate, observational properties and impact of these cold accretion disk winds.

\section{acknowledgements}
We acknowledge support by the Spanish MINECO under grant AYA2017-83216-P. TMD and MAPT acknowledge support via Ram\'on y Cajal Fellowships RYC-2015-18148 and RYC-2015-17854. MAP is funded by the Juan de la Cierva Fellowship Programme (IJCI--2016-30867). DMS acknowledges support from the ERC under the European Union’s Horizon 2020 research and innovation programme (grant agreement No. 715051; Spiders). JvdE and ND are supported by an NWO Vidi grant awarded to ND. FF and DA acknowledge support from the Royal Society International Exchanges \textit{The first step for High-Energy Astrophysics relations between Argentina and UK}. DA and DJKB acknowledge support from the Royal Society. MOA acknowledges support from the Royal Society through Newton International Fellowship program. N.C.S. acknowledge support by the Science and Technology Facilities Council (STFC), and from STFC grant ST/M001326/1. FMV acknowledges support from STFC under grant ST/R000638/1. {\sc molly} software developed by Tom Marsh is acknowledged. 

\bibliographystyle{aasjournal}
\bibliography{/Users/tmd/Dropbox/Libreria} 

\end{document}